\documentclass[twocolumn,showpacs,preprintnumbers,amsmath,amssymb]{revtex4}

\usepackage{graphicx}
\usepackage{dcolumn}
\usepackage{bm}

\begin{document}

\preprint{}

\title{Crossover from dilute-Kondo system to heavy-fermion system}

\author{Hiroshi Watanabe}
 \email{h-watanabe@riken.jp}
\affiliation{%
RIKEN, 2-1, Hirosawa, Wako-shi, Saitama 351-0198, Japan
}%

\author{Masao Ogata}%
\affiliation{%
Department of Physics, University of Tokyo, 7-3-1, Hongo, Bunkyo-ku, Tokyo 113-0033, Japan
}%

\date{\today}

\begin{abstract}
Ground state properties of a Kondo lattice model with random configuration of $f$ electrons are investigated with a variational Monte Carlo method.
We show that the crossover from a dilute-Kondo system to a heavy-fermion system occurs when the density of $c$ and $f$ electrons ($n_c$, $n_f$) become
comparable, $n_f\lesssim n_c$.
In the heavy-fermion region, the correlation between $f$ electrons is strong and the $f$ electrons themselves greatly contribute to the screening 
of other $f$-electron moments.
We propose that the character of Kondo screening changes from ``individual'' to ``collective'' across the crossover.

\end{abstract}

\maketitle

Heavy-fermion materials show various interesting phenomena and have been extensively studied so far.
As is well known, $f$ electrons with a strong localized character play an important role in this system, especially for a large effective mass.
Origins of this large effective mass are considered to be the following factors: 
hybridization between $f$ electrons and other conduction ($c$) electrons, strong Coulomb interaction between $f$ electrons, and a certain number 
of $f$ electrons.
In this letter, we focus on the third factor, namely, relation between the large effective mass and the number of $f$ electrons.

When some magnetic ions with $f$ electrons are replaced by the non-magnetic ions,
e.g., Ce$\rightarrow$La or Yb$\rightarrow$Lu, the system gradually changes from the heavy Fermi liquid (dense limit) to 
the local Fermi liquid (single-impurity limit)~\cite{Sumiyama, Lin, Nakatsuji1}.
However, the details of this crossover are quite different in each compound. 
For example, in Ce$_{1-x}$La$_x$Pb$_3$, the resistivity shows local-Fermi-liquid behavior even at $x=0.2$~\cite{Lin}. 
It is interesting that the single-impurity picture is valid even in a rather ``dense'' region.
On the other hand, the situation is rather complex in Ce$_{1-x}$La$_x$CoIn$_5$~\cite{Nakatsuji1}.
Between the local Fermi liquid and heavy Fermi liquid, 
there is an intermediate region where a temperature scale characterizing the Fermi liquid is greatly suppressed.
It is proposed that the specific heat, magnetic susceptibility, and resistivity can be described by a sum of single-impurity and 
heavy-fermion contributions~\cite{Nakatsuji2, Barzykin}.
This ``two fluid description'' indicates that the system is fluctuating between the local Fermi liquid and heavy Fermi liquid in the intermediate
region.
However, the detailed properties of this region has not been clarified so far.

For this problem, several model Hamiltonians, such as a periodic Anderson model (PAM)~\cite{Yoshimori, Li, Mutou, Grenzebach} 
and a Kondo lattice model (KLM)~\cite{Kaul, Burdin}, have been investigated. 
A crossover from the dilulte-Kondo system to the heavy-fermion system is well described by these studies.
However, the character of Kondo screening and the correlation between $f$ electrons have not been discussed in detail.

In this letter, we study the ground state properties of the KLM with random configurations of $f$ electrons, particularly focusing on the
Kondo screening and the correlation between $f$ electrons.
We show that the crossover from the dilute-Kondo system to the heavy-fermion system occurs when the density of $c$ and $f$ electrons ($n_c$, $n_f$)
become comparable, $n_f\lesssim n_c$.
In the dilute-Kondo region, the $f$ electrons behave independently and the single-impurity picture is valid.
In the heavy-fermion region, the correlation between $f$ electrons becomes strong and the $f$ electrons themselves greatly 
contribute to the screening of other $f$-electron moments.
We show the change in the character of Kondo screening from ``individual'' to ``collective'' across the crossover.
The variational Monte Carlo (VMC) method is used for calculation. This method makes it possible to treat the localized $f$ electrons correctly and to
take into account the effect of randomness beyond the averaged treatment of coherent potential approximation.

We consider the following KLM in a two-dimensional square lattice,
\begin{equation}
 H=\sum_{\bm{k}\sigma}\varepsilon_{\bm{k}}c^{\dagger}_{\bm{k}\sigma}c_{\bm{k}\sigma}
  +J\sum_{i\in \mathrm{A}}\bm{S}_i\cdot\bm{s}_i.\label{Ham}
\end{equation}
where $\varepsilon_{\bm{k}}$ is the energy dispersion of $c$ electrons and $c^{\dagger}_{\bm{k}\sigma} (c_{\bm{k}\sigma})$
is a creation (annihilation) operator of $c$ electrons with momentum $\bm{k}$ and spin $\sigma$.
$\bm{S}_i$ and $\bm{s}_i$ represent the $f$- and $c$-electron spins, namely, 
$\bm{S}_i=\frac{1}{2}\sum_{\sigma\sigma'}f^{\dagger}_{i\sigma}\bm{\sigma}_{\sigma\sigma'}f_{i\sigma'}$ and 
$\bm{s}_i=\frac{1}{2}\sum_{\sigma\sigma'}c^{\dagger}_{i\sigma}\bm{\sigma}_{\sigma\sigma'}c_{i\sigma'}$.
$J(>0)$ denotes the antiferromagnetic exchange coupling between them.
The $f$ electrons are completely localized and treated as localized spins. 
We call the lattice sites with and without $f$ electrons A and B sites, respectively. 
To calculate the physical quantities, we take the average over 10-40 random configurations of A and B sites. 

\begin{figure}[t!]
\begin{center}
\includegraphics[width=8.5cm]{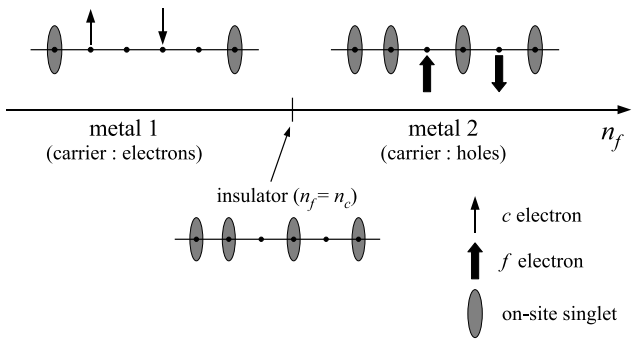}
\caption{\label{fig1}Schematic phase diagram of the KLM in the strong-coupling limit as a function of $n_f$.} 
\end{center}
\end{figure}

First, we consider the strong-coupling limit ($J/t\gg 1$).
We fix the density of $c$ electrons per site ($n_c$) and change the density of $f$ electrons $n_f$ ($0<n_f\leq 1$) on the lattice with 
$N_{\mathrm{s}}$ sites, as shown in Fig.~\ref{fig1}.
In the strong-coupling limit, $c$ electrons at an A site forms an on-site singlet pair with the localized $f$-electron spin. 
When $n_f<n_c$, all localized spins form the on-site singlets and remaining $N_{s}$($n_c-n_f$) $c$ electrons become mobile carriers (called as metal 1). 
On the other hand, when $n_f>n_c$, $c$ electrons are ``exhausted''~\cite{Nozieres} and $N_{s}$($n_f-n_c$) unscreened localized spins exist. 
In this case, when a $c$ electron moves to a site with an unscreened localized spin (unscreened A site), 
the original on-site singlet disappears and another one is newly created.
Thus the hopping of $c$ electrons is described as a motion of on-site singlet among the A sites.
It will cover all the A sites and will dynamically screen all the localized spins.
We can say that this screening is ``collective'' in the sense that $f$ electrons effectively participate the screening of other $f$-electron moments. 
The motion of on-site singlets can be regarded as a motion of mobile holes at the unscreened A sites and it is a different picture with metal 1 (metal 2). 
In the dense limit ($n_f=1$), this system is identical to the $U=\infty$ Hubbard model with a reduced (half) band width~\cite{Lacroix}. 
Thus the metal 2 is considered to be a simple model for heavy-Fermi liquid.
The metal 1 and metal 2 have different characters and separated by the special density of $n_f=n_c$ (see Fig.~\ref{fig1}), 
where the system has a gap of order $J/t$ and becomes insulator. For a finite value of $J/t$, this simple insulating state is not 
expected but it is possible that the character of metallic state changes around $n_f=n_c$~\cite{Kaul, Burdin}. 

For finite $J/t$, we use the VMC method to calculate the physical quantities of the ground state.
We prepare the Gutzwiller-type trial wave function $\left|\Psi \right>$ for a trial state~\cite{Shiba}. 
It consists of a projection operator $\hat{P_f}$ and a one-body part $\left|\Phi \right>$ that is obtained from the solution of 
a certain one-body Hamiltonian.
The specific form is as follows,
\begin{equation}
 \left|\Psi \right> = \hat{P_f}\left|\Phi \right>,
   \;\;\;\; \hat{P_f} = \prod_{i\in \mathrm{A}}\left[ \hat{n}^f_{i\uparrow}(1-\hat{n}^f_{i\downarrow})+\hat{n}^f_{i\downarrow}
                        (1-\hat{n}^f_{i\uparrow})\right].
\end{equation}
$\hat{P_f}$ keeps the $f$-electron number of each A site exactly one. Since the $f$ electrons must be treated as localized spins in
the KLM, this constraint is necessary for the accurate calculation~\cite{Watanabe1}. 
The one-body part $\left|\Phi \right>$ is obtained by diagonalizing the $U=0$ PAM in a real-space representation. 
The explicit form is as follows,

\begin{figure}[t!]
\begin{center}
\includegraphics[width=8.0cm]{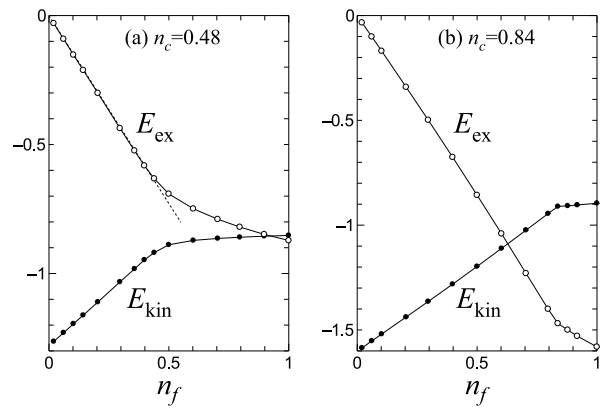}
\caption{\label{fig2}$n_f$ dependence of a kinetic energy $E_{\mathrm{kin}}$ and an exchange energy $E_{\mathrm{ex}}$
for $n_c=0.48$ and $n_c=0.84$ ($J/t=3.0$).} 
\end{center}
\end{figure}

\begin{equation}
H'=\sum_{\left<i,j\right>\sigma}\tilde{t}_{ij\sigma}c^{\dagger}_{i\sigma}c_{j\sigma}+\sum_{i,\sigma}\tilde{E_f}_{i\sigma}f^{\dagger}_{i\sigma}f_{i\sigma}
  +\sum_{i,\sigma}\tilde{V}_{i\sigma}(f^{\dagger}_{i\sigma}c_{i\sigma}+\mathrm{h.c.}). 
\end{equation}

Here, $\tilde{t}_{ij\sigma}$ are transfer integrals from site $j$ to $i$ with spin $\sigma$.
$\tilde{V}_{i\sigma}$ and $\tilde{E_f}_{i\sigma}$ are ``effective'' hybridization and $f$-electron energy level for site $i$ with spin $\sigma$.
To avoid the occupancy of $f$ electrons at B sites, we set $\tilde{E_f}_{i\sigma}=\infty$ 
(or a sufficiently large value in the actual calculation) for $i\in \mathrm{B}$.
$\tilde{t}_{ij\sigma}$, $\tilde{V}_{i\sigma}$ and $\tilde{E_f}_{i\sigma}$ $(i\in \mathrm{A})$ are variational parameters and optimized so as to minimize 
the variational energy.
Here, we set $\tilde{t}_{ij\sigma}=\tilde{t}$, $\tilde{V}_{i\sigma}=\tilde{V}$ and $\tilde{E_f}_{i\sigma}=\tilde{E_f}$ for simplicity.  
Then we calculate physical quantities with the optimized wave function.
To obtain the expectation values, we take 10-40 patterns of random configurations and average them.
Although a large number of different configurations are necessary, this calculation procedure will take account of the effect of randomness 
more precisely than the coherent potential approximation that is often used for disordered systems.

We perform a VMC calculation for a square lattice with $\varepsilon_{\bm{k}}=-2t(\cos k_x+\cos k_y)$.
$t=1$ is set to be a unit of energy in the following.
The size dependence of results have been checked from 10$\times$10 to 16$\times$16.
In the system with 10$\times$10 sites, we have found some finite-size effects at $n_f\lesssim0.16$~\cite{Watanabe2} 
but they disappear in the system with more than 14$\times$14 sites. 
Since the 10$\times$10 system is not sufficient for describing the dilute region, we show the results of 14$\times$14 in the following.

\begin{figure}[t!]
\begin{center}
\includegraphics[width=7.0cm]{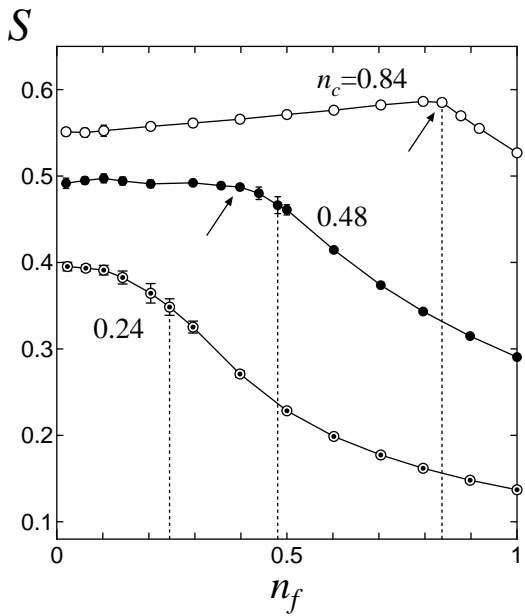}
\caption{\label{fig3}$n_f$ dependence of strength of on-site screening $S$ for $n_c=0.24$, 0.48 and 0.84 ($J/t=3.0$). Dotted lines represent
$n_f=0.24$, 0.48 and 0.84. Arrows indicate the crossover points.} 
\end{center}
\end{figure}

First, Fig.~\ref{fig2} shows the $n_f$ dependence of kinetic energy $E_{\mathrm{kin}}$ and exchange energy $E_{\mathrm{ex}}$ that
are expectation values of the first and second term of right-hand side of Eq.~(\ref{Ham}). 
$n_f\rightarrow0$ corresponds to the dilute limit and $n_f=1$ corresponds to the usual KLM.
For the case with $n_c=0.48$ (Fig.~\ref{fig2}(a)), $E_{\mathrm{ex}}$ shows almost linear behavior until $n_f\simeq0.40$, namely, 
the exchange energy per $f$ electron ($E_{\mathrm{ex}}/N_f$) is almost unchanged even with increasing the number of $f$ electrons. 
It means that the additional $f$ electrons do not affect the screening of other $f$ electrons so much. 
We can say that the $f$ electrons in this region behave independently and the single-impurity picture is valid until $n_f\simeq0.40$.
Because of the finite value of $J/t$, this boundary is slightly different from that expected in the strong-coupling limit, 0.48.
The effect of finite $J/t$ is discussed later. For the case with $n_c=0.84$ (Fig.~\ref{fig2}(b)), 
as can be seen by the abrupt changes in $E_{\mathrm{kin}}$ and $E_{\mathrm{ex}}$, 
the single-impurity picture is valid until $n_f\simeq0.84$. 
It is almost the same with the strong-coupling limit.

Next, we calculate the following quantity,
\begin{equation}
S=-\frac{E_{\mathrm{ex}}}{JN_f}=-\frac{1}{N_f}\left<\sum_{i\in \mathrm{A}}\bm{S}_i\cdot\bm{s}_i\right>_{\mathrm{rnd}}. \label{Eex}
\end{equation}
where $\left<\cdots\right>_{\mathrm{rnd}}$ denotes the random average.
This quantity is always positive and corresponds to the strength of on-site screening per $f$ electron.
We fix the value of $n_c$ as 0.24, 0.48 and 0.84 and show the $n_f$ dependence of $S$ for $J/t=3.0$ in Fig.~\ref{fig3}. 
For $n_c=0.84$, as $n_f$ increases, $S$ gradually increases and rapidly decreases at $n_f\simeq n_c=0.84$.
This crossover at $n_f\simeq n_c$ is consistent with the strong-coupling limit, and thus corresponds to the crossover 
from the dilute-Kondo system to the heavy-Fermion system.
As shown later, the character of screening changes from individual to collective at this density.
For $n_c=0.48$, on the other hand, the crossover occurs at $n_f\simeq0.40$, slightly away from $n_c$.
We consider that this is because the screening clouds have a finite spread for a finite value of $J/t$.
In the strong-coupling limit, the screening clouds are completely localized at each site and have no overlap.
In this case, the boundary between the dilute-Kondo system (individual screening) and heavy-fermion system (collective screening) is exactly at $n_f=n_c$.
However, the screening can be collective if the neighboring screening clouds overlap and interact with each other, even though $n_f<n_c$.
For $n_c=0.48$, the spread of screening cloud will exceed the lattice spacing and as a result, the region of heavy-fermion system 
extends to $n_f<n_c$.
For $n_c=0.24$, the change of $S$ becomes more gradual and we cannot determine the clear crossover point from these data.  
We conclude that the more $n_c$ decreases, the more the screening cloud spreads and the crossover becomes ambiguous.

\begin{figure}[t!]
\begin{center}
\includegraphics[width=8.5cm]{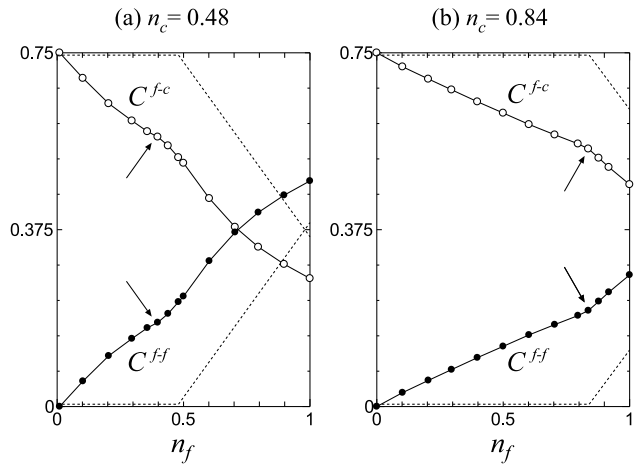}
\caption{\label{fig4}$n_f$ dependence of $C^{f-c}$ and $C^{f-f}$ for (a) $n_c=0.48$ and (b) $n_c=0.84$ ($J/t=3.0$). Dotted lines represent the
strong-coupling limit. Arrows indicate the crossover points.} 
\end{center}
\end{figure}

To observe the character of screening more clearly, we study the correlation between spins.
For each $f$-electron spin $\bm{S}_i$, the following relation is satisfied,
\begin{equation}
C_i^{f-c}+C_i^{f-f}\equiv-\sum_{j}\bm{S}_i\cdot\bm{s}_j-\sum_{j\neq i}\bm{S}_i\cdot\bm{S}_j=+\frac{3}{4}, \label{C}
\end{equation}
because $\bm{S}_i$ forms singlet pairing with the rest of all spins in a paramagnetic state.
We calculate $C^{f-c}=\bigl<C_i^{f-c}\bigr>$ and $C^{f-f}=\bigl<C_i^{f-f}\bigr>$, where $\bigl<\cdots\bigr>$ denotes an average over both sites and 
random configurations.
$C^{f-c}$ ($C^{f-f}$) indicates how much the $c$ ($f$) electrons are concerning with the screening.
Figure~\ref{fig4} shows the $n_f$ dependence of these quantities for $n_c=0.48$ and 0.84.
In the single-impurity limit ($n_f\rightarrow0$), screening is performed only by $c$ electrons and $C^{f-c}\rightarrow$ 3/4 (0.75) and 
$C^{f-f}\rightarrow0$. 
As $n_f$ increases, $f$ electrons begins to participate the screening and $C^{f-f}$ increases while $C^{f-c}$ decreases.  
At a certain point around $n_f\simeq n_c$ (indicated by arrows in Fig.~\ref{fig4}), an abrupt change is observed. 
It corresponds to the crossover point shown in Fig.~\ref{fig3} and represents the crossover from the dilute-Kondo system to the heavy-fermion system.
In the heavy-fermion system, $f$ electrons themselves are greatly participating the screening of other $f$-electron moments and 
therefore the correlation between $f$ electrons is much stronger than in the dilute-Kondo system. 
We consider that this strong correlation between $f$ electrons is an origin of large effective mass in the heavy-fermion system.

Here, we compare our result with other theoretical works. Mean-field calculation by Kaul and Vojta for 20$\times$20 square lattice~\cite{Kaul} has shown 
the $n_f$ dependence of $T^*=\left<b_r^2\right>/D$, where $2b_r=J\left<c^{\dagger}_rf_r\right>$ denotes auxiliary fields and $D$ denotes a half 
band width.
$T^*$ is almost the same quantity with $S$ in our result and also shows the crossover around $n_f\simeq n_c$.
Burdin and Fulde have calculated $T_0/T_{\mathrm{K}}$ using a coherent potential approximation and a dynamical mean-field theory approach~\cite{Burdin},
where $T_0$ denotes a coherence (Fermi liquid) temperature scale and $T_{\mathrm{K}}$ denotes a Kondo temperature.
$T_0/T_{\mathrm{K}}$ is an important ratio and related to the effective mass as $T_0/T_{\mathrm{K}}=m^*_{\mathrm{SIKM}}/m^*_{\mathrm{KLM}}$,
where SIKM denotes single-impurity Kondo model ($n_f\rightarrow0$ limit of KLM).
Their result have shown the rapid decrease of $T_0/T_{\mathrm{K}}$, namely, the enhancement of effective mass around $n_f\simeq n_c$. 
It is consistent with our result showing that the crossover from the dilute-Kondo system to the heavy-fermion system occurs around $n_f\lesssim n_c$.
The enhancement of effective mass for $n_f>n_c$ is also shown in the PAM by Pruschke \textit{et al}~\cite{Pruschke}.

Finally, we refer to the relation between our result and the experiments. 
Yb$_{1-x}$Lu$_{x}$Al$_3$ shows clear deviation from the single-impurity model up to $x\sim0.7$ $(n_f\sim0.3)$ and the heavy-fermion picture is 
valid in this region~\cite{Bauer}.
The estimation of $n_c\sim0.5$ in this compound~\cite{Cornelius} does not contradict our result.
The systematic experiment of Yb$X$Cu$_4$ has shown that the compounds with large $n_c$ ($X$=Ag, Tl) are well described by the single-impurity
model, while the compounds with small $n_c$ ($X$=Mg, Zn, Cs) are not~\cite{Lawrence}.
This result is reasonable at least qualitatively if the former correspond to the dilute-Kondo system ($n_f<n_c$) and the latter correspond to 
the heavy-fermion system ($n_f>n_c$).
In Ce$_{1-x}$La$_x$CoIn$_5$, an intermediate region with both the single-impurity and the heavy-fermion contributions are observed~\cite{Nakatsuji2}.
This indicates that the crossover from the dilute-Kondo system to the heavy-fermion system is rather ambiguous and the system is fluctuating 
between them.
In our result, the ambiguous crossover appears in the case of small $J/t$ (not shown in this paper) and small $n_c$ because the screening cloud is 
fairly extended in this case.
We expect a smaller $J/t$ and/or a smaller $n_c$ for Ce$_{1-x}$La$_x$CoIn$_5$, compared to the materials with clear crossover, 
such as Ce$_{1-x}$La$_x$Pb$_3$.

In this paper, we have studied the ground state properties of the KLM with random configuration of $f$ electrons.
The VMC method in real space is used for calculations.
We have found that the crossover from the dilute-Kondo system to the heavy-fermion system occurs around $n_f\lesssim n_c$.
In the dilute-Kondo region, the $f$ electrons behave independently and the single-impurity picture is valid.
In the heavy-fermion region, on the other hand, the correlation between $f$ electrons is strong and the $f$ electrons themselves participate the 
screening of other $f$-electron moments.
It is considered to be an origin of large effective mass.
We propose that the screening character changes from ``individual'' to ``collective'' across the crossover. 

\begin{acknowledgments}
The authors thank Y. Yanase and S. Yunoki for useful discussions.
This work is supported by a Grant-in-Aid for Scientific Research on Innovative Areas ``Heavy Electrons'' (20102002) from the
Ministry of Education, Culture, Sports, Science and Technology, Japan and also by a Next Generation Supercomputing Project, Nanoscience Program, 
MEXT. The computation in this work has been done using the RIKEN Cluster of Clusters (RICC) facility and the facilities of the Supercomputer Center,
Institute for Solid State Physics, University of Tokyo.
\end{acknowledgments}


\begin{thebibliography}{99} 
\bibitem{Sumiyama} A. Sumiyama, Y. Oda, H. Nagano, Y. \^{O}nuki, K. Shibutani and T. Komatsubara,  
                   J. Phys Soc. Jpn. \textbf{55}, 1294 (1986).
\bibitem{Lin} C. L. Lin, A. Wallash, J. E. Crow, T. Mihalisin, and P. Schlottmann, Phys. Rev. Lett. \textbf{58}, 1232 (1987).
\bibitem{Nakatsuji1} S. Nakatsuji, S. Yeo, L. Balicas, Z. Fisk, P. Schlottmann, P. G. Pagliuso, N. O. Moreno, J. L. Sarrao, and J. D. Thompson, 
                    Phys. Rev. Lett. \textbf{89}, 106402 (2002).   
\bibitem{Nakatsuji2} S. Nakatsuji, D. Pines, and Z. Fisk, Phys. Rev. Lett. \textbf{92}, 016401 (2004).
\bibitem{Barzykin} V. Barzykin, Phys. Rev. B \textbf{73}, 094455 (2006).
\bibitem{Yoshimori} A. Yoshimori and H. Kasai, Solid State Commun. \textbf{58}, 259 (1986).
\bibitem{Li} Z. Z. Li and Y. Qiu, Phys. Rev. B \textbf{43}, 12906 (1991).
\bibitem{Mutou} T. Mutou, Phys. Rev. B \textbf{64}, 245102 (2001).
\bibitem{Grenzebach} C. Grenzebach, F. B. Anders, G. Czycholl, and T. Pruschke, Phys. Rev. B \textbf{77}, 115125 (2008).
\bibitem{Kaul} R. K. Kaul and M. Vojta, Phys. Rev. B \textbf{75}, 132407 (2007).
\bibitem{Burdin} S. Burdin and P. Fulde, Phys. Rev. B \textbf{76} 104425 (2007).
\bibitem{Nozieres} P. Nozi\`{e}res, Eur. Phys. J. B \textbf{6}, 447 (1998).
\bibitem{Lacroix} C. Lacroix, Solid State Commun. \textbf{54}, 991 (1985).
\bibitem{Shiba} H. Shiba and P. Fazekas, Prog. Theor. Phys. Suppl. \textbf{101}, 403 (1990).     
\bibitem{Watanabe1} H. Watanabe and M. Ogata, Phys. Rev. Lett. \textbf{99}, 136401 (2007).
\bibitem{Watanabe2} H. Watanabe and M. Ogata, to be appeared in J. Phys.
\bibitem{Bauer} E. D. Bauer, C. H. Booth, J. M. Lawrence, M. F. Hundley, J. L. Sarrao, J. D. Thompson, P. S. Riseborough, and T. Ebihara,
                Phys. Rev. B \textbf{69}, 125102 (2004).  
\bibitem{Cornelius} A. L. Cornelius, J. M. Lawrence, T. Ebihara, P. S. Riseborough, C. H. Booth, M. F. Hundley, P. G. Pagliuso, J.L. Sarrao,
                    J. D. Thompson, M. H. Jung, A. H. Lacerda, and G. H. Kwei, Phys. Rev. Lett. \textbf{88}, 117201 (2002).               
\bibitem{Lawrence} J. M. Lawrence, P. S. Riseborough, C. H. Booth, J. L. Sarrao, J. D. Thompson, and R. Osborn, Phys. Rev. B \textbf{63}, 054427 (2001). 
\bibitem{Pruschke} T. Pruschke, R. Bulla, and M. Jarrell, Phys. Rev. B \textbf{61}, 12799 (2000).

\end{thebibliography}
\end{document}